\*

# Unsupervised Spectral Demosaicing with Lightweight Spectral Attention Networks

Kai Feng, Yongqiang Zhao, Seong G. Kong, and Haijin Zeng

*Abstract*—This paper presents a deep learning-based spectral demosaicing technique trained in an unsupervised manner. Many existing deep learning-based techniques relying on supervised learning with synthetic images, often underperform on real-world images especially when the number of spectral bands increases. According to the characteristics of the spectral mosaic image, this paper proposes a mosaic loss function, the corresponding model structure, a transformation strategy, and an early stopping strategy, which form a complete unsupervised spectral demosaicing framework. A challenge in real-world spectral demosaicing is inconsistency between the model parameters and the computational resources of the imager. We reduce the complexity and parameters of the spectral attention module by dividing the spectral attention tensor into spectral attention matrices in the spatial dimension and spectral attention vector in the channel dimension, which is more suitable for unsupervised framework. This paper also presents *Mosaic2*5, a real 25-band hyperspectral mosaic image dataset of various objects, illuminations, and materials for benchmarking. Extensive experiments on synthetic and real-world datasets demonstrate that the proposed method outperforms conventional unsupervised methods in terms of spatial distortion suppression, spectral fidelity, robustness, and computational cost.

*Index Terms*—spectral demosaicing, unsupervised learning, spectral imaging, spectral attention networks

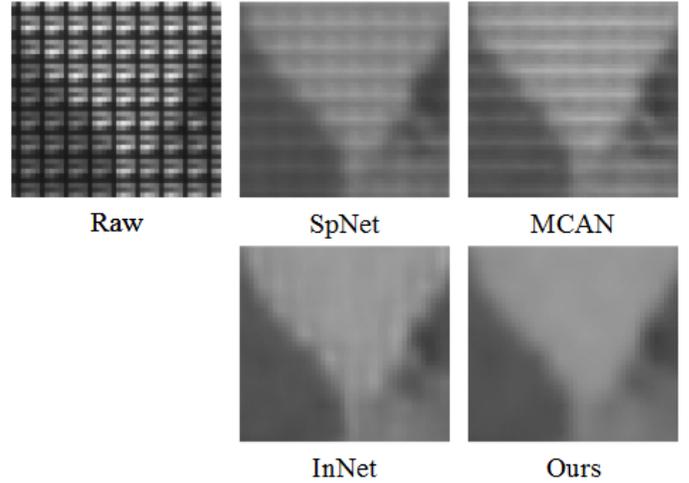

**Fig. 1.** Demosaicing results at 841nm by supervised deep demosaicing networks SpNet [8], MCAN[9], InNet [10], and the proposed unsupervised deep demosaicing network on a real spectral mosaic image of a leaf captured by a 25-band SFA-based spectral imager.

## I. INTRODUCTION

Spectral imaging acquires both spectral as well as spatial information simultaneously, which have been useful for applications including vaccine detection [1], clinical monitoring[2], and object tracking[3]–[5]. However, a large device size and slow imaging speed of spectral cameras often limit practical applicability of spectral imaging. Snapshot spectral cameras based on compact micro spectral filter arrays (SFAs) [6], [7] have been utilized to overcome the limitation of conventional spectral imaging sensors.

Demosaicing is essential to reconstruct a full spectral image from the signals measured using an SFA-based imager. Many existing demosaicing techniques can be divided into unsupervised and supervised methods. Most unsupervised methods rely on interpolation [11]–[16], and conventional model optimization [17]–[19]. These methods are robust yet underperform in the high-frequency region. Supervised methods [8]–[10], [20]–[23] mostly rely on paired mosaic images and fully defined cubes to train deep neural networks. Current fully defined image cubes used in spectral demosaicing are mostly synthesized by the dataset images taken by other non-snapshot spectral cameras. As spectral resolution increases, spatial resolution of a single band image decreases, and therefore demosaicing networks rely more on spectral correlations. However, the spectral correlations in synthetic images are likely inaccurate [24]. Because the real image is the weighted integral of the continuous waveband images, we synthesize it as the weighted sum of the discrete waveband images in the dataset. The spectral response of each waveband image in the dataset is a narrow band function treated as a unit impulse function [25]. Such reasons may cause the supervised methods trained by synthetic images usually fail in real-world

This work was supported by the National Natural Science Foundation of China (NSFC) under Grant 61771391, the Key R & D plan of Shaanxi Province 2020ZDLGY07-11, the Science, Technology and Innovation Commission of Shenzhen Municipality under Grants JCYJ20170815162956949 and JCYJ20180306171146740, the National Research Foundation of Korea under Grant NRF-2016R1D1A1B01008522, the Fundamental Research Funds for the Central Universities. *(Corresponding author: Yongqiang Zhao.)*

Kai Feng, Yongqiang Zhao are with the School of Automation, Northwestern Polytechnical University, Xi'an 710072, China (e-mail: 2018100620@mail.nwpu.edu.cn; zhaoyq@nwpu.edu.cn).

Seong G. Kong is with the Department of Computer Engineering, Sejong University, Seoul 05006, Korea (e-mail: skong@sejong.edu).

Haijin Zeng is with the Image Processing and Interpretation imec research group at Ghent University, 9000 Ghent, Belgium (e-mail: Haijin.Zeng@UGent.be)

cases. Fig. 1 shows demosaicing results of several representative supervised spectral demosaicing techniques trained using synthetic images. The supervised methods produce global spatial distortions or artifact in the edge. Higher-quality datasets and better synthesis methods can alleviate the problem, which is not trivial to achieve. This paper solves the stated problem by training neural network directly on real mosaic images in an unsupervised manner.

Inspired by unsupervised equivariant imaging (EI) [26], we propose an Unsupervised Spectral Demosaicing (USD) framework including training method, model structure, transformation enhancement strategy, and early stopping strategy. With the USD framework, we do not need to prepare paired synthetic data and only use the real mosaic images to train deep networks. Specifically, we first design an unsupervised observation loss function based on the sampling matrix of spectral mosaic image; To prevent the model from collapsing, we propose a necessary condition on the model structure: the interpolation branch; Considering that the periodic sampling characteristic of the SFA pattern makes the number of intuitive shifting transformations in EI limited, we then propose a mixed transformation strategy including shifting, rotation, resizing, and flipping to enhance the performance of USD; To further detect the model overfitting in USD, we also propose a self-evaluation index, which measures the similarity between the periodic down sampled sub-images of each band image. Utilizing this index, we can automatically stop training at the appropriate epoch.

Another challenge with deep spectral demosaicing is inconsistency between the model parameters and computational resources of the imager. Demosaicing is essential in SFA-based spectral imaging that ideally needs to be implemented into the device [27]. In practice, however, snapshot spectral imaging devices have limited memory spaces to store a large number of parameters [28]. Spectral attention is an effective technique in supervised spectral demosaicing [9], but with a large amount of parameters and fails in USD. This paper presents a lightweight spectral attention (LSA) module with fewer parameters by dividing the spectral attention tensor into the spectral attention matrixes among spatial dimension and the spectral attention vector among channel dimension. The proposed LSA not only reduces 99.8% of the parameters of the spectral attention module, but also improves the performance of USD.

To verify the validity of the proposed scheme, we built *Mosaic25*, a real-world 25-band hyperspectral mosaic image dataset containing various objects, illuminations, and materials. This dataset is expected to serve as a benchmark data in the future.

Extensive experiments were conducted on synthetic data and a real-world 25-band hyperspectral mosaic image dataset to compare the performances of the proposed technique with the state-of-the-art spectral demosaicing methods. Experiment results show that our proposed method outperforms conventional supervised and unsupervised techniques in terms of spatial distortion suppression, spectral fidelity, robustness, and computational cost.

Main contributions of the proposed demosaicing technique are:
1) Presents an unsupervised spectral demosaicing (USD) framework including loss function, model structure, transformation strategy, and early stopping strategy.
2) Proposes a lightweight spectral attention module that significantly reduces the number of parameters and improves the performance of USD.
3) Develops *Mosaic25*, a real-world 25-band hyperspectral mosaic image dataset containing various objects, illuminations, and materials.

## II. SPECTRAL DEMOSAICING TECHNIQUES

This section introduces the spectral demosaicing problem, the proposed unsupervised spectral demosaicing framework, together with the designed lightweight spectral attention module.

### A. Problem Overview

A common spectral filter array is periodic. The mosaic sampling can be written as:

$$Y = \sum_{b=1}^{B} X_b \odot M_b \quad (1)$$

where $X \in \mathbb{R}^{H \times W \times B}$ represents the fully defined spectral cube, with $H$, $W$, and $B$ as the height, width, and band numbers of the image, respectively, $X_b \in \mathbb{R}^{H \times W}$, $\forall b = 1,...,B$ represents the $b$-th band image, $M \in \mathbb{R}^{H \times W \times B}$ denotes the mosaic sampling mask of SFA, $M_b$ is the $b$-th mask, and $Y \in \mathbb{R}^{H \times W}$ denotes the observed spectral mosaic image.

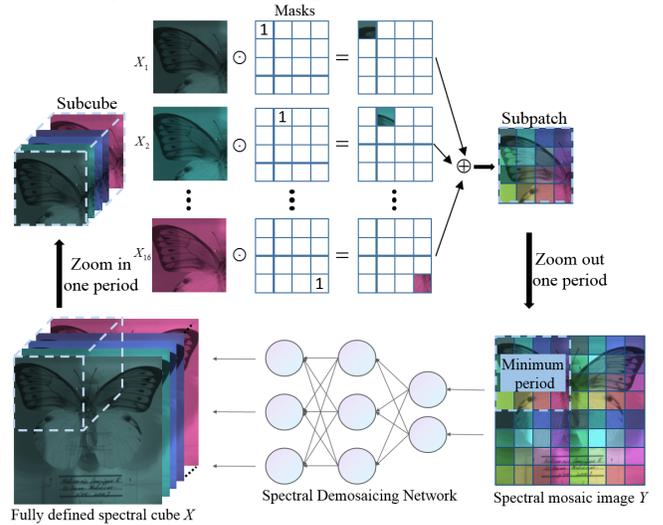

**Fig. 2.** A schematic diagram of the sampling and demosaicing process of a spectral mosaic image. A 4×4 SFA pattern is shown as an example.

The mosaic sampling is irreversible. The goal is to construct a module, e.g., a deep demosaicing network, whose input is $Y$ and the output is the estimated fully defined cube $\hat{X}$, as shown in the bottom part of Fig. 2. Previous supervised studies minimize the loss between the estimated $\hat{X}$ and ground truth $X$ to optimize the parameters $\theta$ of the network. The final optimized parameters $\hat{\theta}$ can be formulated as:

$$\hat{\theta} = \arg\min_{\theta} l(\hat{X}, X) = \arg\min_{\theta} l(F(Y,\theta), X) \quad (2)$$

where $l(\cdot)$ is the loss function for the fully defined spectral cube. In most supervised studies, $X$ is synthesized using data from other non-snapshot spectral cameras, so the paired $Y$-$X$ can be easily obtained to make the above training process work.

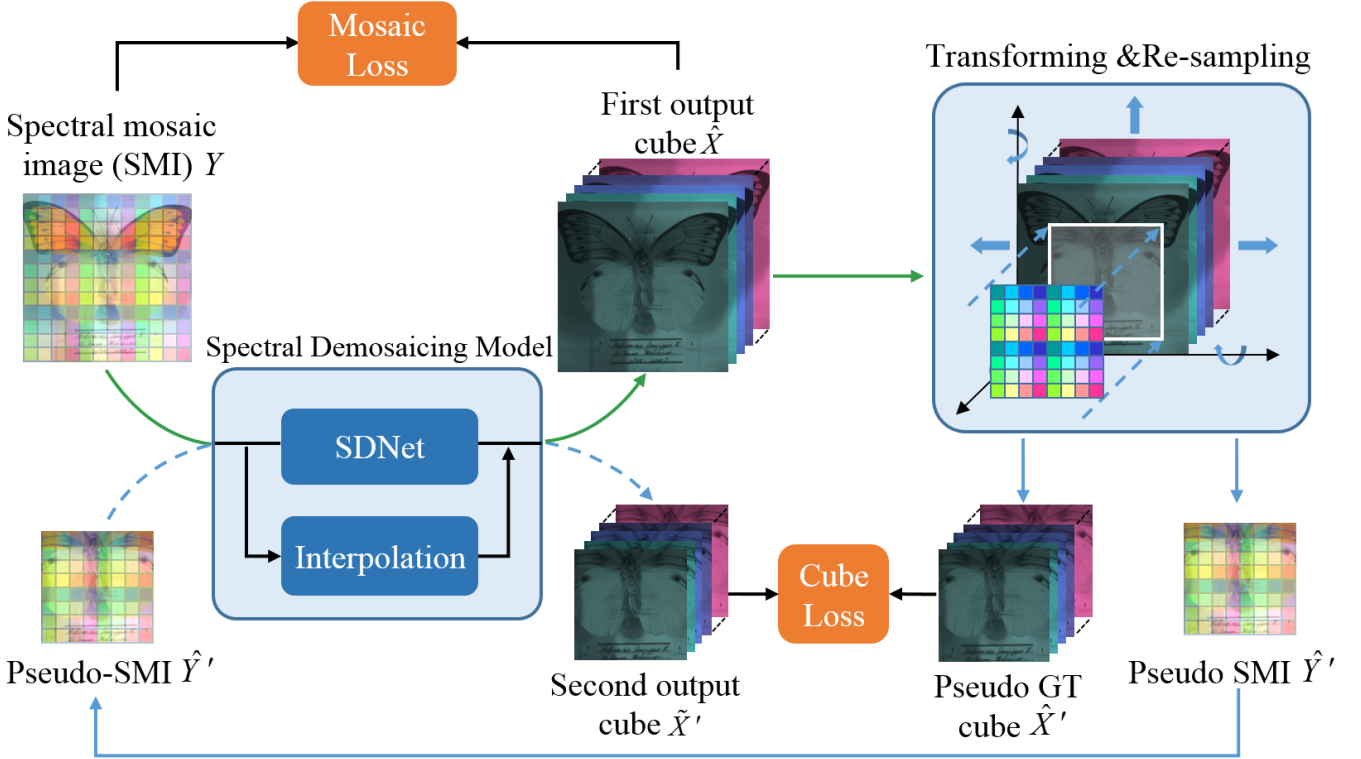

**Fig. 3.** A schematic diagram of the proposed unsupervised spectral demosaicing (USD) framework. We feed the real spectral mosaic image (SMI) $Y$ to spectral demosaicing model and obtain the first demosaiced cube $\hat{X}$. Then we transform and re-sample this cube to obtain the pseudo fully defined ground truth cube $\hat{X}'$ and SMI $\hat{Y}'$, and feed $\hat{Y}'$ to the previous model to obtain the second demosaiced cube $\tilde{X}'$. Finally, with the mosaic loss between $Y$ and $\hat{X}$, the cube loss between $\tilde{X}'$ and $\hat{X}'$, we can train spectral demosaicing model in an unsupervised manner.

### B. Unsupervised Spectral Demosaicing Framework

*Motivation.* In general, it is not easy to obtain the real fully-defined radiance spectral cube $X$. In CFA (Color Filter Array) demosaicing, researchers benefit from well-established commercial cameras to capture ground truth, i.e., fully defined cube, by shifting the sensor chip [29], which is expensive and imperfect for the SFA imager [30]. Also inspired by equivariant imaging [26], we shift the demosaiced cube output by the network, then re-sample this demosaiced cube with the known sampling matrix and feed the new spectral mosaic image to the previous network. The new demosaiced cube should be the same as the previous demosaiced cube if this network demosaics perfectly. Based on the above motivations, we design an unsupervised spectral demosaicing framework. Fig. 3 shows the overall schematic diagram of the USD framework.

*Basic training method.* Here we describe the detailed training process of USD. First, we feed the real mosaic image $Y$ that has no paired spectral cube $X$ to demosaicing network:

$$\hat{X} = F(Y;\theta) \quad (3)$$

where $\hat{X}$ is the first output cube of the demosaicing network $F$.

Then we shift the first output of the network $\hat{X}$. Considering the mosaic sampling matrix is periodic, we randomly shift the mosaic sampling matrix only at the range of one period, which is different from the original global shift operation in the inpainting task [26], the shifted cube $\hat{X}'$ can be represented as:

$$\hat{X}' = T_{shift}(\hat{X}, i, j) \quad (4)$$

where $i \in [1, r_1]$ and $j \in [1, r_2]$ represent the number of pixels to shift in the horizontal and vertical directions of the spatial dimension, $r_1, r_2$ are the pattern size of SFA.

Then we take the $\hat{X}'$ as pseudo fully defined cube and sample $\hat{X}'$ to obtain the new pseudo spectral mosaic image (SMI) $\hat{Y}'$:

$$\hat{Y}' = \sum_{b=1}^{B} \hat{X}'_b \odot M_b \quad (5)$$

Now, we obtain the pseudo paired $\hat{Y}'$-$\hat{X}'$. We feed $\hat{Y}'$ to the demosaicing network used before to obtain the second output cube $\tilde{X}'$:

$$\tilde{X}' = F(\hat{Y}';\theta) \quad (6)$$

Then, we can solve the following optimization problem for spectral demosaicing without the real fully defined spectral cube $X$:

$$\begin{aligned}\hat{\theta} &= \arg\min_{\theta}\left\{l_{cube}\left(\tilde{X}',\hat{X}'\right)+\alpha l_{mosaic}\left(\hat{X},Y\right)\right\}\\&=\arg\min_{\theta}\left\{l_{cube}\left(F(\hat{Y}';\theta),\hat{X}'\right)+\alpha l_{mosaic}\left(F(Y;\theta),Y\right)\right\}\end{aligned} \quad (7)$$

where $l_{cube}(\cdot)$ is the equivariant loss between shifted cubes, we use Charbonnier $L_1$ loss [31] to implement $l_{cube}(\cdot)$, $l_{mosaic}(\cdot)$ is the proposed mosaic loss function, hyper-parameter $\alpha$ is used to control the weight of mosaic loss.

**Mosaic loss.** In the spectral demosaicing task, the known pixels in a raw spectral mosaic image are valuable and reliable. We hope the demosaiced cube can be sampled back to the raw spectral mosaic image, which is also a necessary condition for the establishment of unsupervised imaging, therefore, we represent the mosaic loss as follows:

$$l_{mosaic}\left(\hat{X},Y\right)=\frac{1}{H\times W}\left\|\sum_{b=1}^{B}\hat{X}_b\odot M_b - Y\right\|_1 \quad (8)$$

where $M$ is the mosaic sampling mask of SFA, $H$, $W$ are the height and width of the raw spectral mosaic image, $\|\cdot\|_1$ is the Charbonnier $L_1$ loss.

**Model Structure.** From (7), there exists a model collapse when the network only repeats the input mosaic image at the spectral dimension as output. This model can also make the loss value the lowest but without demosaicing. So we add the interpolated cube that is non-mosaic to the output of the network, as shown in Fig. 4. This interpolated cube is the main low-frequency component. We encourage the network to predict the residual high-frequency component to avoid the above model collapse. This paper uses classical weighted bilinear (WB) interpolation algorithm [16] which interpolates each sparse band independently so that the interpolated cube will not have any periodic mosaic characteristic.

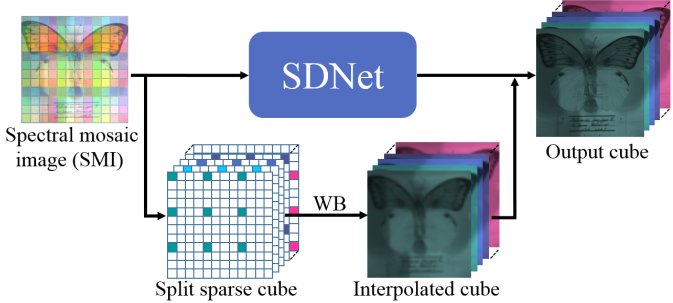

**Fig. 4.** A schematic diagram of the proposed model structure for USD framework, where WB means the weighted interpolation.

**Mixed Transformation Strategy.** According to the conclusion in EI [26], the more the number of transformations means the better performance of equivariant learning. Meanwhile, the number of valid shift transformations in demosaicing is little, which is just $r_1 \times r_2$, where $r_1$, $r_2$ are the size of the SFA pattern. So, we propose a mixed transformation enhancement strategy that contains shift, flip, rotation, and resize operations to enrich the number of transformations, as shown in Fig. 3.

**Early Stopping Strategy.** In supervised training, we can detect overfitting, stop the training procedure, and select a good model by computing the error between the output of the network and the ground truth on validation dataset. In an unsupervised training with no ground truth, however, finding an early stopping criterion is not easy. So, we study overfitting of proposed unsupervised spectral demosaicing training. We found that when the unsupervised spectral demosaicing model overfits, there will be periodic distortion in the space of the output cube, as shown in Fig. 5. Therefore, we propose a self-evaluation index to measure the degree of periodic distortion in each band image of output cube.

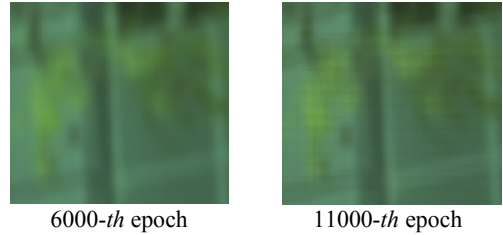

6000-*th* epoch      11000-*th* epoch

**Fig. 5.** The pseudo color results of the demosaicing model under USD training of different training epochs. With the increase in the number of training epochs, the model appears overfitting, which is manifested as periodic distortion in space.

Since we found that the period of distortion is consistent with the SFA pattern size, we first rearrange the $b$-th band image $x_b$ of output cube using Inverse Pixel Shuffle (IPS) technology [32] into $r_1 r_2$ sub-images, then we calculate the variance between the global mean of these sub-images:

$$v_b = Var(Mean(IPS(x_b))) \quad (9)$$

Considering this periodic distortion exists in each band image, the self-evaluation index (SEI) of the whole cube is defined as:

$$SEI = \frac{1}{B}\sum_{b=1}^{B}v_b \quad (10)$$

where $B$ is the number of bands, in a non-redundancy SFA pattern $B = r_1 \times r_2$.

Therefore, we can detect whether the model overfits by supervising the SEI change curve rather than observing the output image of every training epoch manually.

Algorithm 1 shows the pseudo code of the USD algorithm.

---

**Algorithm 1** Pseudo code of the USD algorithm in a PyTorch-like style.

*# SDModel is the deep spectral demosaicing model with an interpolation branch*

while (1):
    for y in minibatchs: *# y is one minibatch that contains some mosaic images*
        x = DMModel(y)
        x_trans = random_transform(x) *# this function randomly selects the transformation from the shift, flip, rotation, and resize operations.*
        y_trans = mosaic_sampling(x_trans) *# this function samples the cube x_trans using the mosaic SFA pattern*
        x_trans1 = SDModel(y_trans)
        loss = CubeLoss(x_trans1, x_trans)
            + MosaicLoss(x, y)

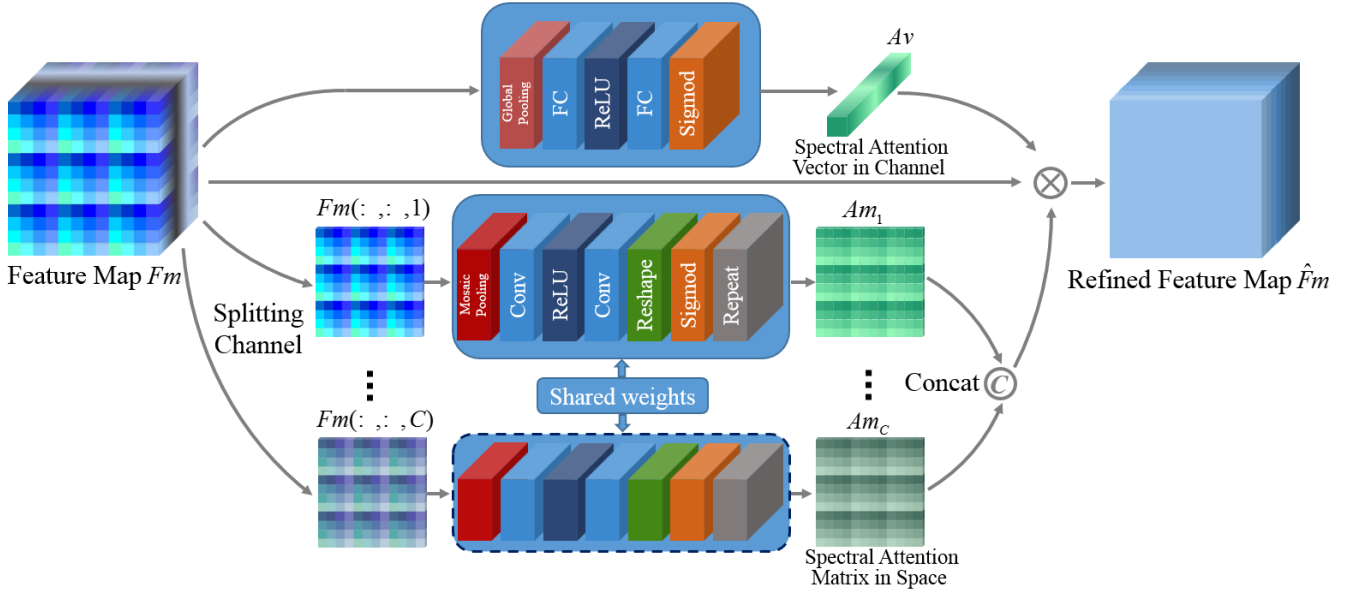

**Fig. 6.** A schematic diagram of the proposed lightweight spectral attention model structure.

```
    loss.backward()
    update(SDModel.params)
    sei = SEI_compute(SDModel, y_val) # y_val are evaluation
spectral mosaic images
    if sei > sei_max:
        break # overfitting happens, stop the training
```

### C. Lightweight Spectral Attention Module

**Motivation.** The spectral attention module has been proven effective in supervised spectral demosaicing [9]. The spectral correlation is distributed on the space and channel of deep feature maps. The spectral attention module learns a 3D attention weight tensor for adjusting the spectral feature map. This way introduces a large number of weight parameters, so we call it heavyweight spectral module (**HSA**), and we found it failed in USD. More details can be found in Section III.C. We propose a lightweight spectral attention (**LSA**) module that splits the attention weight tensor into spectral attention 2D matrixes in space and spectral attention 1D vector in channel, as shown in Fig. 6, aiming to reduce the number of learning parameters of the spectral module and make it useful in USD.

**Spectral attention in space.** We first obtain the spectral attention matrix for each channel of the feature map. The network architecture of spectral attention in space is like that of HSA in [9], but the input to this network is just a single channel feature map. Each channel of the feature map shares the same attention network to obtain the dedicated spectral attention matrix, which can reduce the number of parameters significantly. This process can be expressed as:

$$Am_i = f_{sas}(Fm(:,:,i), \theta_{sas}), i \in \{1, \cdots, C\} \quad (11)$$

where $Am_i$ is the attention matrix of the $i$-th channel of feature map $Fm$, $f_{sas}(\cdot)$ is the function of spectral attention in space, $\theta_{sas}$ are the learning weights of spectral attention in space. $C$ is the channel number of the feature map.

**Spectral attention in channel.** Then we refer to the channel attention in [33] to generate the spectral attention vector in channel $Av$ as follows:

$$Av = f_{sac}(Fm, \theta_{sac}) \quad (12)$$

where $f_{sac}(\cdot)$ is the function of spectral attention in channel, $\theta_{sac}$ are the learning weights of spectral attention in channel.

Finally, we multiply the initial feature map $Fm$ by the attention matrixes $Am$ and attention vector $Av$ to obtain the refined feature map $\hat{Fm}$ as follows:

$$\hat{Fm} = Fm \cdot Am \cdot Av \quad (13)$$

**Discussion of the number of parameters.** The core network architecture of LSA and HSA are all squeeze-excitation modes [33], so the parameters can be easily calculated from the number of input sizes and the reduction ratio $d$.

For one feature map with $C$ channels in spectral demosaicing networks of the $r_1 \times r_2$ SFA pattern, the number of parameters of its corresponding HSA is $2r_1^2 r_2^2 C^2 / d$.

The numbers of parameters of the corresponding spectral attention in the space and in the channel are $2r_1^2 r_2^2 / d$ and $2C^2 / d$, respectively

The total number of parameters of LSA is:

$$2r_1^2 r_2^2 / d + 2C^2 / d = 2(r_1^2 r_2^2 + C^2) / d \quad (14)$$

Typically, $C$ in deep neural networks is selected as a relatively large number such as 64, the SFA pattern size $r_1 \times r_2$ in this paper is 5×5. Therefore, the design of LSA can reduces about 99.8% parameters compared to HSA.

### III. EXPERIMENT RESULTS

#### A. Implementation Details

We replace heavyweight spectral attention (HSA) with the proposed lightweight spectral attention (LSA) on the basis of MCAN [9] as the model for subsequent experiments. We validate our algorithm in a challenging 25-band ($r_1 = r_2 = 5$) snapshot multispectral camera. The setup of the input layer, output layer are modified according to the 5×5 pattern. We set the reduction ratio $d=4$ in LSA, and the weight hyper-parameter $\alpha =1$ in loss.

*B. Datasets and Metrics*

In the experiments with synthetic data, we use hyperspectral images dataset generated by the Interdisciplinary Computational Vision Lab (ICVL) [34] to simulate snapshot multispectral mosaic images, which can validate the performance of our demosaicing algorithm qualitatively and quantitatively. We select 40 images as the training dataset and 10 images as the testing dataset. We sampled the original cube into a 25-band cube using spectral sensitive functions provided by the manufacturer IMEC. We use PSNR, SSIM [35], SAM [36], and ERGAS [37] as the performance evaluation metrics in the experiments with synthetic data.

In the experiments with real data, we propose *Mosaic25*, a real-world 25-band hyperspectral mosaic image dataset of various objects, illuminations, and materials. The images in our dataset are captured by one IMEC 25-band snapshot spectral camera which spectrum ranges from 600nm~900nm. We selected 40 regions of interest from these mosaic images as a training dataset. Examples of these image patches are shown in Fig. 8. Then we additionally provide another 17 full images with size of 2045×1080 as a testing dataset in a variety of objects and lighting conditions. The nature of the scenes ranges from indoor environments to outdoor environments. Our dataset is expected to serve as a common benchmark for spectral demosaicing. We use the no-reference spectral image evaluation metric proposed in [38] to show the performance quantitatively in the experiments with real data.

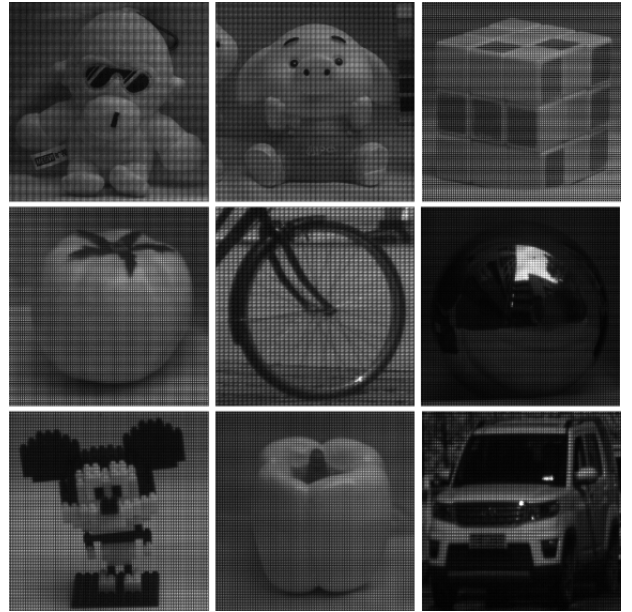

**Fig. 8.** Training examples of the proposed *Mossaic25* Dataset, captured by a real IMEC 25-band snapshot spectral camera. Our dataset covers various objects, illuminations, and materials.

We randomly crop one patch from the training dataset to form a training batch. The spatial size of the patch is 100×100. We set the learning rate as $1×10^{-4}$.

*C. Ablation Study*

**The significance of the proposed unsupervised spectral demosaicing model structure.** To verify the significance of the proposed model structure in USD, we train one model without and one model with an interpolation branch on the ICVL training dataset in the proposed unsupervised training manner and plot their PSNR values on the ICVL test dataset as a function of training epochs in Fig. 7. The model without interpolation branch does not work, but the model with an interpolation branch performs very well illustrating the significance of the interpolation branch in the USD model structure.

**Effectiveness of unsupervised early stopping strategy.** To verify the effectiveness of the self-evaluation index (SEI) in unsupervised early stopping, we plot the PSNR value and SEI value of our model on the ICVL testing dataset as a function of training epochs in Fig. 9. As the number of epochs grow to 6500, the PSNR begins to decrease, indicating the overfitting occurs. At the same time, the SEI value begins to rise and fluctuate in a wide range. Therefore, we can use the SEI value to detect whether the USD model overfits, and select an appropriate model. For the experiments with synthetic data on the ICVL dataset, we set up the maximum SEI as $2.1×10^{-7}$. For the experiments with real data on the Mosaic25 dataset, we set up the maximum SEI as $1×10^{-6}$.

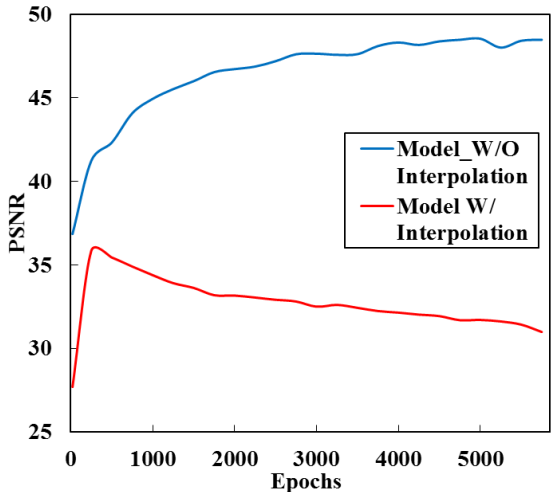

**Fig. 7.** PSNR values of the model with interpolation branch and the model without interpolation at different epochs on ICVL dataset.

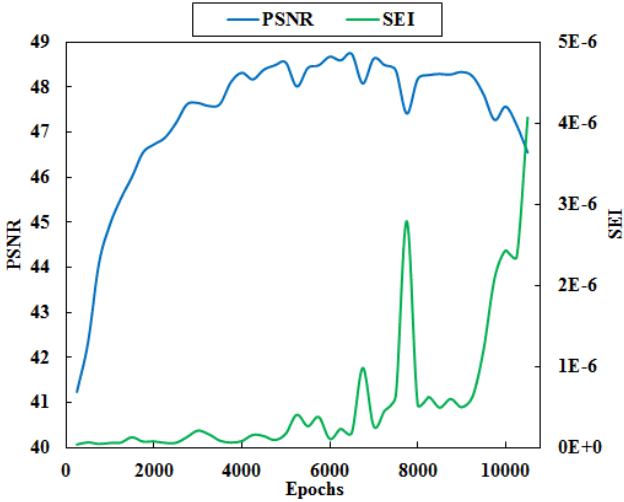

**Fig. 9.** PSNR and SEI value of proposed method at different epochs on ICVL dataset. The decrease in PSNR is in line with the rise in SEI overall.

**Comparison with different transformation strategies.** To verify the effectiveness of the proposed mixed transformation strategy, we compare the performance of the model with different training strategies. Specifically, three models are exploited here. The first model is denoted *ShiftTran*, which only uses the shifting transformation strategy. The second model is denoted *MixTrans*, which is trained by the proposed mixed transformations strategy. The last model is trained by the classical supervised training strategy, denoted *Supervised*.

Table I shows the average results obtained by *ShiftTran*, *MixTrans*, and *Supervised* over all testing images in the ICVL dataset. From this table, we can observe mixed transformations strategy is more effective than the shifting transformation strategy and close to the classical supervised training strategy.

TABLE I
AVERAGE PERFORMANCE OF DIFFERENT TRAINING STRATEGIES OVER ICVL TESTING DATASET

|  | Methods | PSNR↑ | SSIM↑ | SAM↓ | ERGAS↓ |
|---|---|---|---|---|---|
| USD | ShiftTran | 46.67 | 0.9959 | 0.99 | 2.99 |
|  | MixTrans | 48.89 | 0.9977 | 0.86 | 2.24 |
|  | Supervised | 50.19 | 0.9983 | 0.75 | 1.93 |

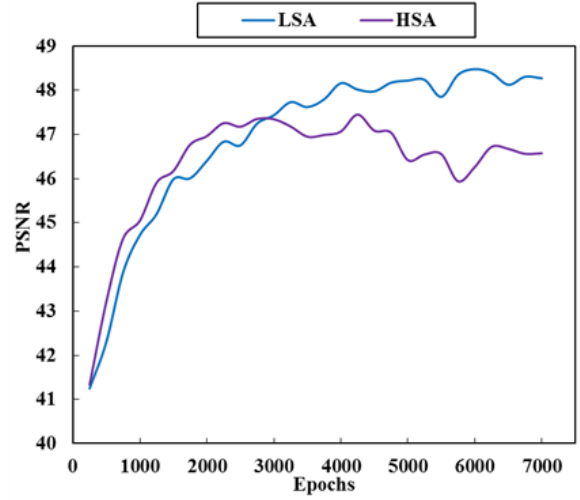

**Fig. 10.** PSNR values of the model with HSA and the model with LSA at different training epochs on ICVL dataset.

**Comparison with heavyweight spectral attention module.** To verify the efficiency of the proposed lightweight spectral attention module, we train the base model without the spectral attention module labelled as *Base*, the model with heavyweight spectral attention labelled as *HSA* to compare with our model labelled as *LSA* here. Both models are trained under the USD framework.

TABLE II
AVERAGE PERFORMANCE OVER THE ICVL TESTING DATASET AND THE NUMBER OF MILLION PARAMETERS OF THE BASE MODEL WITH DIFFERENT SPECTRAL MODULE UNDER THE USD FRAMEWORK

| Model | PSNR↑ | SSIM↑ | SAM↓ | ERGAS↓ | Param(M) |
|---|---|---|---|---|---|
| Base | 48.59 | 0.9975 | 0.87 | 2.33 | 0.411 |
| HSA | 46.67 | 0.9964 | 1.08 | 2.85 | 3.166 |
| LSA | 48.89 | 0.9977 | 0.86 | 2.24 | 0.416 |

Table II shows the average results over the ICVL testing dataset and the number of million parameters of *Base*, *HSA*, and *LSA*. The parameters of *HSA* are an order of magnitude more than those of the *Base*, but its performance is worse, which shows that *HSA* fails under USD framework. Compared to *HSA*, *LSA* not only reduces the parameters, but also outperforms *Base*.

TABLE III
DEMOSAICING PERFORMANCE (PSNR/SSIM/SAM/ERGAS) AT THREE TYPICAL SCENES AND THE AVERAGE PERFORMANCE OVER ALL TEST SCENES FOR DIFFERENT METHODS IN ICVL DATASET

| Methods | Nachal | | | | Grf | | | | Bgu | | | | Average of all | | | |
|---|---|---|---|---|---|---|---|---|---|---|---|---|---|---|---|---|
|  | PSNR | SSIM | SAM | ERGAS | PSNR | SSIM | SAM | ERGAS | PSNR | SSIM | SAM | ERGAS | PSNR | SSIM | SAM | ERGAS |
| WB | 36.42 | 0.964 | 1.35 | 6.25 | 37.44 | 0.989 | 1.33 | 7.36 | 35.18 | 0.988 | 1.60 | 6.67 | 39.12 | 0.982 | 1.47 | 6.52 |
| ItSD | 35.96 | 0.950 | 1.55 | 7.35 | 39.46 | 0.993 | 1.44 | 6.01 | 37.51 | 0.993 | 1.82 | 5.28 | 40.80 | 0.986 | 1.61 | 5.65 |
| PPID | 37.87 | 0.971 | 1.19 | 5.59 | 40.11 | 0.994 | 1.11 | 5.42 | 38.33 | 0.994 | 1.31 | 4.62 | 41.92 | 0.990 | 1.23 | 4.81 |
| Ours | 45.22 | 0.994 | 0.75 | 2.41 | 46.75 | 0.999 | 0.81 | 2.56 | 46.73 | 0.999 | 0.72 | 1.79 | 48.89 | 0.998 | 0.86 | 2.24 |
| Ideal Value | ↑ | ↑1 | ↓0 | ↓0 | ↑ | ↑1 | ↓0 | ↓0 | ↑ | ↑1 | ↓0 | ↓0 | ↑ | ↑1 | ↓0 | ↓0 |

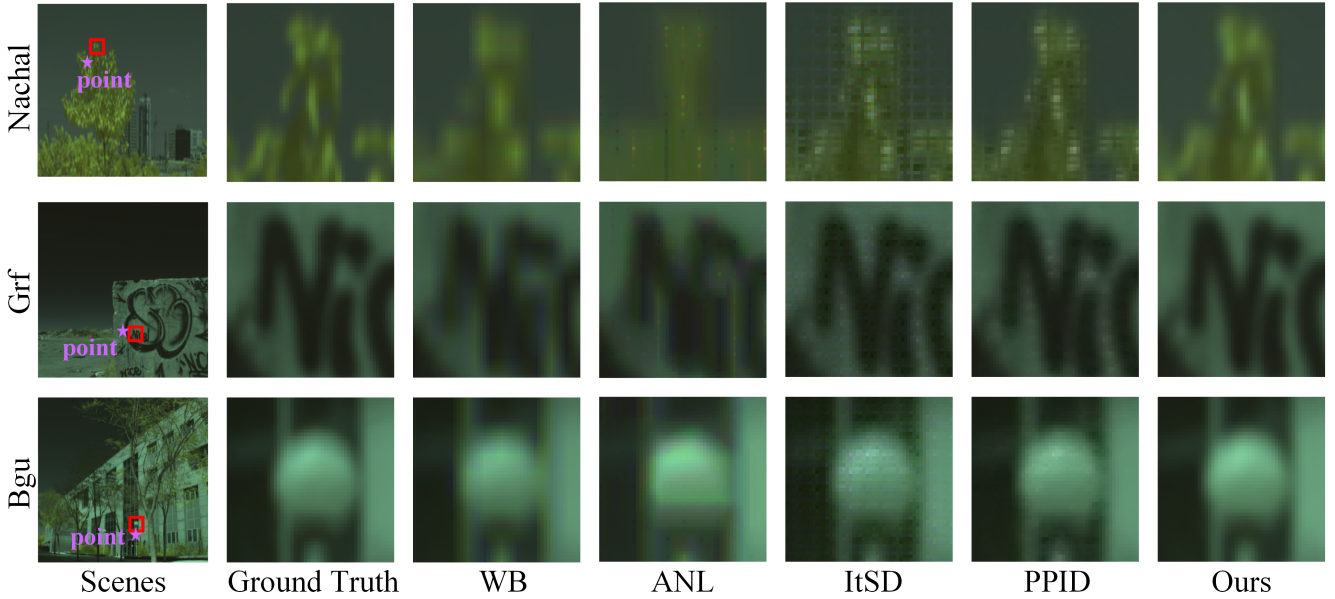

**Fig. 11.** The local details of the pseudo color ground truth and demosaicing results of the conventional unsupervised methods and the proposed method for three ICVL scenes.

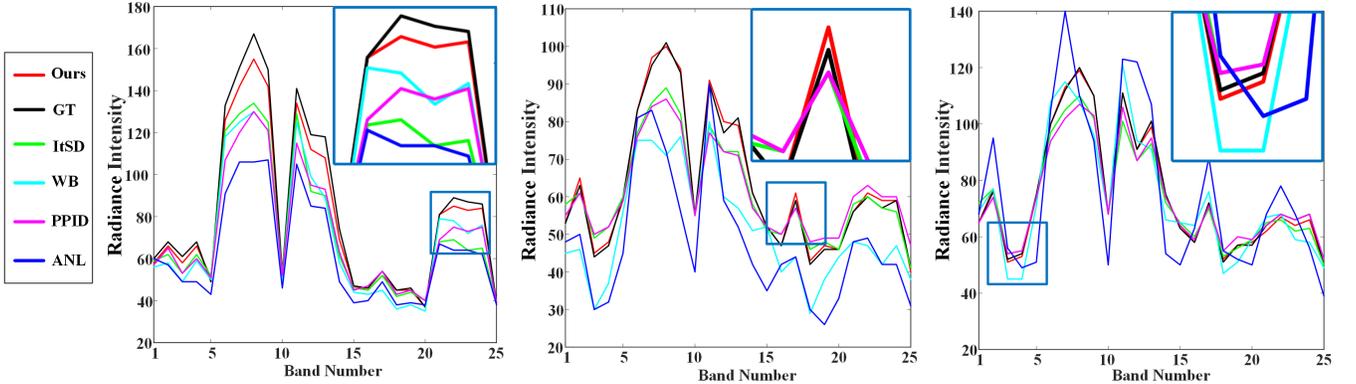

**Fig. 12.** The radiance spectrum of the purple marked points in Fig. 11 of ground truth and demosaicing results of the conventional unsupervised methods and the proposed method.

Fig. 10 shows PSNR values of *HSA* and *LSA* at different training epochs on ICVL dataset. Overfitting occurs early in *HSA*, and the maximum value of PSNR is relatively low, which further proves that the proposed lightweight spectral attention module is more suitable for USD framework.

*D. Comparisons on Synthetic Data*

We then evaluate the proposed model (the base model with our lightweight spectral module) under the proposed USD training framework on the synthetic ICVL dataset. We conducted experiments to compare with state-of-the-art unsupervised and supervised methods.

**Comparison methods.** The unsupervised spectral demosaicing methods for comparison are WB [16], ANL [18], ItSD [12], and PPID [13]. ANL is a non-local based method, testing on the full ICVL image is very time-consuming, so we just evaluate ANL qualitatively on the 250×250 patch, which will slightly decrease its performance. The supervised comparison methods include representative deep neural network based methods: SpNet [8], MCAN [9], and InNet [10] all retrained on the ICVL dataset.

**Performance comparison on unsupervised methods.** Table III lists the performance at three representative scenes and average performance over all ICVL testing scenes of three comparison methods and our method. The proposed method significantly outperforms other competing methods on all evaluation metrics. Figs. 11 and 12 show the (23, 13, 5)-band results in pseudo color and the radiance spectrum of the representative points of the three scenes of all methods. Figs. 13 and 14 show (1, 22, 12)-band results in pseudo color and the radiance spectrum of another three scenes. Our proposed method outperforms the other techniques for comparison, in both recovery of spatial textures and spectral fidelity.

**Performance comparison on supervised methods.** Table IV lists the average quantitative indexes of supervised methods and our unsupervised method on ICVL dataset. In terms of PSNR and SSIM reflecting the spatial recovery performance, our method just has a small gap with the best MCAN, and is better than InNet. In terms of SAM reflecting the spectral recovery performance, our method is only inferior to the best InNet. Our method is competitive even compared with the supervised method on synthetic data.

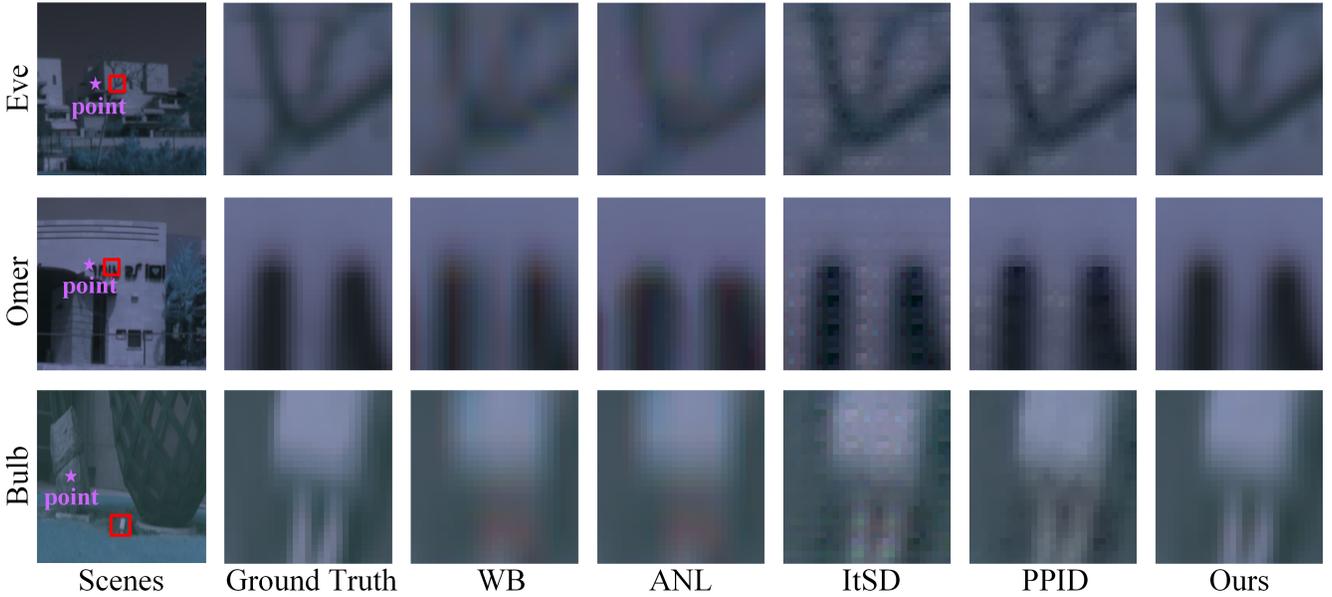

**Fig. 13.** The local details of the pseudo color ground truth and demosaicing results of conventional unsupervised methods and the proposed method for three ICVL scenes.

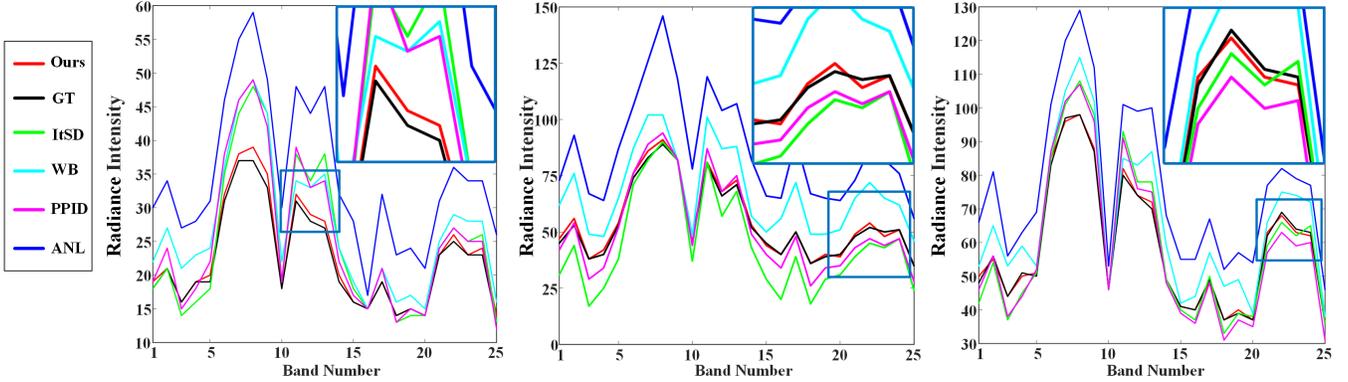

**Fig. 14.** The radiance spectrum of the purple marked points in Fig. 13 of ground truth and demosaicing results of conventional unsupervised methods and the proposed method.

*E. Comparisons of Real Data*

We then evaluate our method in the case that training and testing data are all from the proposed real-world Mosaic25 dataset without ground-truth data, as compared with existing unsupervised methods and representative supervised methods.

TABLE IV
AVERAGE PERFORMANCE OF REPRESENTATIVE SUPERVISED METHODS AND OUR UNSUPERVISED METHOD OVER ALL ICVL TESTING DATASET (THE BEST RESULT IS BOLD IN BLACK, AND THE SECOND BEST RESULT IS UNDERLINED IN BLUE)

| Methods | PSNR ↑ | SSIM ↑ | SAM ↓ | ERGAS ↓ |
|---|---|---|---|---|
| SpNet | 49.26 | 0.9979 | 0.871 | 2.11 |
| MCAN | **49.93** | **0.9982** | 0.865 | **1.97** |
| InNet | 48.03 | 0.9972 | **0.764** | 2.46 |
| Ours | 48.89 | 0.9977 | 0.861 | 2.24 |

TABLE V
NO-REFERENCE PERFORMANCE OF ALL COMPETING METHODS AT THREE TYPICAL SCENES AND AVERAGE PERFORMANCE OVER REAL MOSAIC25 TESTING DATASET

| Methods | Card | Gate1 | Gate2 | Average |
|---|---|---|---|---|
| WB | 21.94 | 22.18 | 21.06 | 21.72 |
| ItSD | 20.33 | 19.60 | 19.95 | 19.91 |
| PPID | 20.96 | 19.60 | 19.61 | 19.82 |
| Ours | **19.46** | **19.57** | **19.41** | **19.72** |

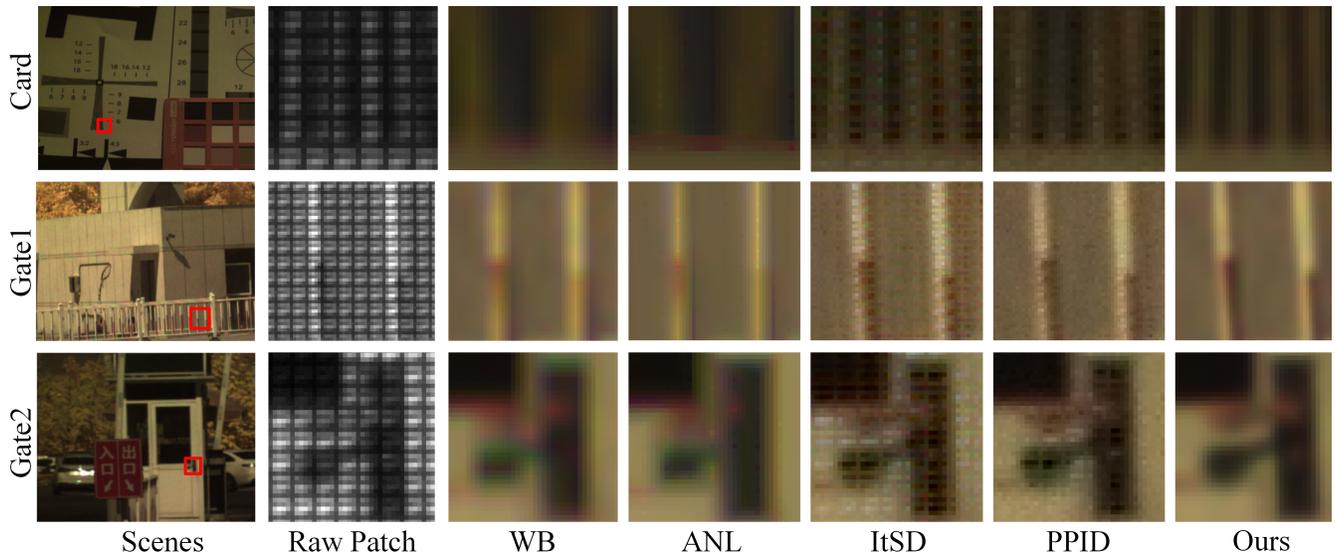

**Fig. 15.** The local details of real-world raw mosaic and pseudo color demosaicing results of the unsupervised methods for comparison and our method for three real scenes from Mosaic25.

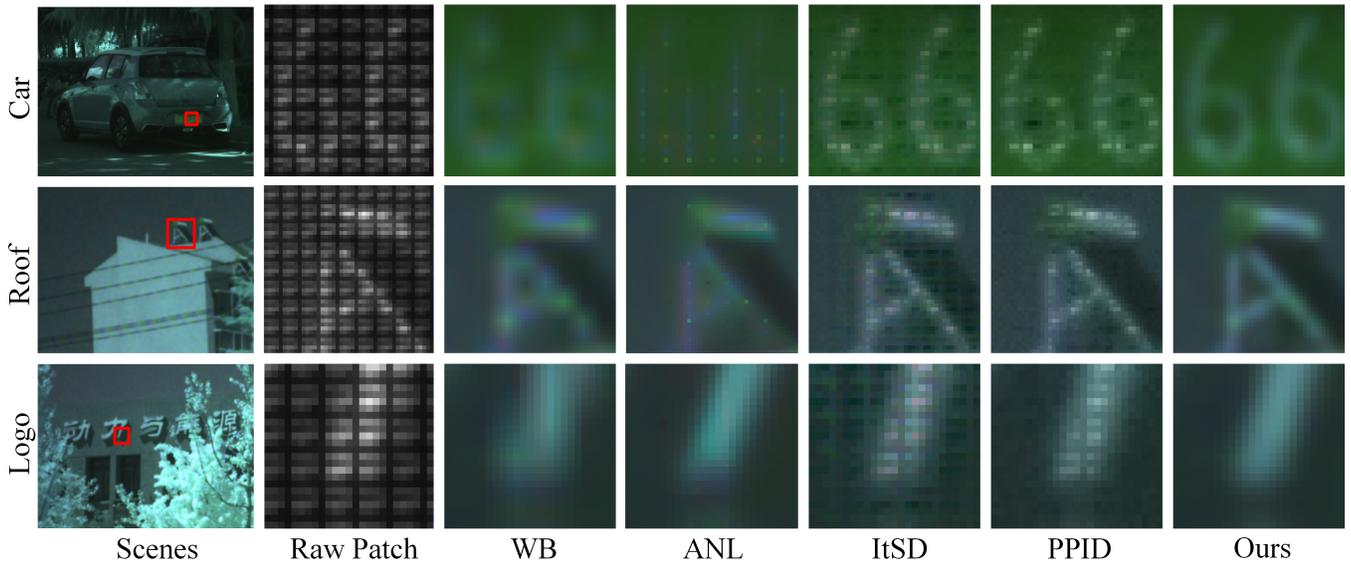

**Fig. 16.** The local details of real-world raw mosaic and pseudo color demosaicing results of the unsupervised methods for comparison and our method for three real scenes from Mosaic25.

**Performance comparison on unsupervised methods.** We compare the performance of all competing unsupervised methods on the Mosaic25 real-world dataset. Table V shows the no-reference spectral image evaluation [38] results of three representative scenes and the average result of overall testing dataset. Our method significantly outperforms other methods. Figs. 12 and 13 visually compares different pseudo color results on six representative scenes obtained by all competing methods. Our method achieves better results than other methods and is robust on real data.

**Performance comparison on supervised methods.** Although we have explained in the Introduction section that the supervised method trained on synthetic data does not perform well in real data, we show the more results of the supervised method and our proposed unsupervised method in Fig. 17. The SpNet and MCAN show strong periodic distortion, and SpNet has color distortion, indicating the loss of spectral information. The InNet shows artifacts in high-frequency areas. Our method outperforms these representative supervised methods in terms of spectral fidelity, spatial details and distortion suppression.

*F. Running Times and Computational Cost*

Demosaicing is a fundamental step for an SFA-based camera, so the running time and computational cost of demosaicing method is an important evaluation dimension. We test the running times of all the comparison method and our method when demosaicing one 1300×1390 mosaic image on one Intel(R) Xeon(R) Gold 6240 CPU and NVIDIA RTX 2080Ti

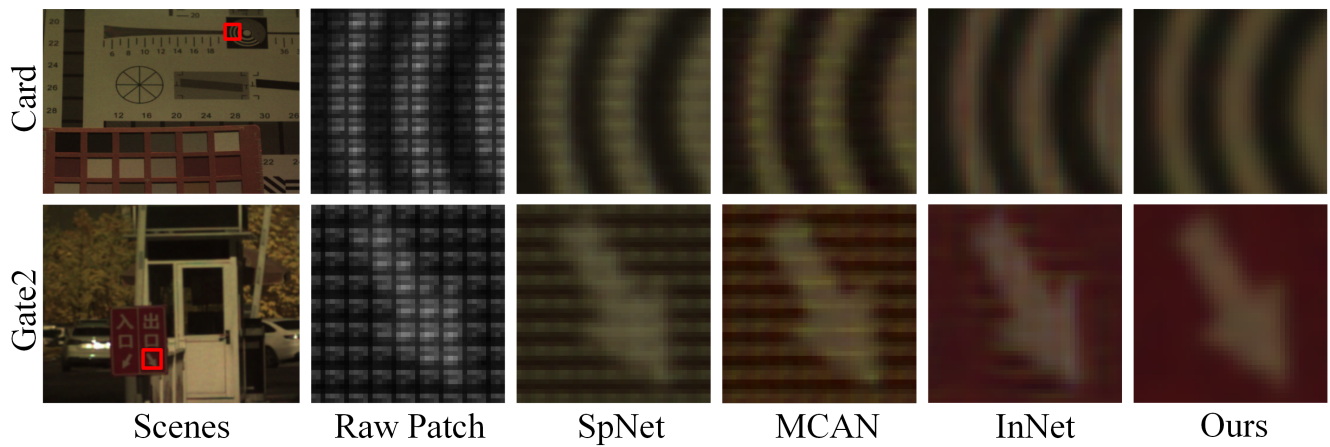

**Fig. 17.** The local details of real-world raw mosaic and pseudo color demosaicing results of the supervised methods and our method for two real scenes from Mosaic25.

GPU. The number of million parameters and TFLOPs of deep models are also test. In Table VI, the running time are averaged over 50 times and the first image are not included to avoid the impact of machine startup, the running time of the ANL method is obtained by reasoning the running time on a patch of 250×250 size, InNet is based on the memory-consuming 3D convolution, and cannot get complete results at one time, so we test the image in blocks for InNet. The CPU running speed of our method is the second only to the fastest WB method. The GPU running speed of SpNet, MCAN, and our proposed method are similar, and can reach more than 200 frames per second, but the parameters and TFLOPs of our proposed method are smaller. Our proposed method is more comprehensive and suitable for deployment on real devices.

TABLE VI
RUNNING TIMES, NUMBER OF MILLION PARAMETERS, AND TFLOPS OF ALL COMPETING METHODS ON ONE 25-BAND SPECTRAL MOSAIC IMAGE OF SIZE 1300×1390 (THE BEST RESULT IS BOLD IN BLACK, AND THE SECOND BEST RESULT IS UNDERLINED IN BLUE. THE '\' MEANS THIS TERM IS NOT COUNTED IN NON-DEEP LEARNING METHODS)

| Platform | Methods | Running time(s) | #Parameters (Million) | TFLOPs |
|---|---|---|---|---|
| CPU | WB | **0.8** | \ | \ |
| | ItSD | 73.1 | \ | \ |
| | PPID | 35.1 | \ | \ |
| | NL | 572400.0 | \ | \ |
| | Ours | 4.7 | 0.4 | 0.5 |
| GPU | SpNet | 0.004 | 1.2 | 2.3 |
| | MCAN | **0.003** | 3.2 | **0.5** |
| | InNet | 7.0 | 0.9 | 39.0 |
| | Ours | 0.004 | **0.4** | **0.5** |

## IV. CONCLUSION

This paper presents an unsupervised spectral demosaicing framework including the basic training method, model structure, enhancement strategy, early stopping strategy, and a lightweight attention network module. Together with the effective and efficient series of solutions proposed above, the proposed spectral demosaicing method holds promise for future deployment into practical devices and the creation of a real spectral video dataset with high spectral-spatial-temporal resolution. Future research extension of this work includes unsupervised joint denoising and demosaicing [39], [40], no-reference image quality assessment [41], super-resolution [42], and object detection [43], [44].

REFERENCES

[1] A. Yin, Y. Wang, Y. Chen, K. Zeng, H. Zhang, and J. Mao, "SSAPN: Spectral-Spatial Anomaly Perception Network for Unsupervised Vaccine Detection," *IEEE Transactions on Industrial Informatics*, pp. 1–12, 2022, doi: 10.1109/TII.2022.3195168.

[2] J. R. Bauer, A. A. Bruins, J. Y. Hardeberg, and R. M. Verdaasdonk, "A Spectral Filter Array Camera for Clinical Monitoring and Diagnosis: Proof of Concept for Skin Oxygenation Imaging," *J. Imaging*, vol. 5, no. 8, p. 66, Jul. 2019, doi: 10.3390/jimaging5080066.

[3] F. Xiong, J. Zhou, and Y. Qian, "Material Based Object Tracking in Hyperspectral Videos," *IEEE Transactions on Image Processing*, vol. 29, pp. 3719–3733, 2020, doi: 10.1109/TIP.2020.2965302.

[4] L. Chen et al., "Object tracking in hyperspectral-oriented video with fast spatial-spectral features," *Remote Sensing*, vol. 13, no. 10, p. 1922, 2021.

[5] L. Chen, Y. Zhao, J. C.-W. Chan, and S. G. Kong, "Histograms of oriented mosaic gradients for snapshot spectral image description," *ISPRS Journal of Photogrammetry and Remote Sensing*, vol. 183, pp. 79–93, Jan. 2022, doi: 10.1016/j.isprsjprs.2021.10.018.

[6] J. Jia, K. J. Barnard, and K. Hirakawa, "Fourier Spectral Filter Array for Optimal Multispectral Imaging," *IEEE Transactions on Image Processing*, vol. 25, no. 4, pp. 1530–1543, Apr. 2016, doi: 10.1109/TIP.2016.2523683.

[7] Y. Monno, S. Kikuchi, M. Tanaka, and M. Okutomi, "A Practical One-Shot Multispectral Imaging System Using a Single Image Sensor," *IEEE Transactions on Image Processing*, vol. 24, no. 10, pp. 3048–3059, Oct. 2015, doi: 10.1109/TIP.2015.2436342.

[8] T. A. Habtegebrial, G. Reis, and D. Stricker, "Deep Convolutional Networks For Snapshot Hypercpectral Demosaicking," in *2019 10th Workshop on Hyperspectral Imaging and Signal Processing: Evolution in Remote Sensing (WHISPERS)*, Sep. 2019, pp. 1–5. doi: 10.1109/WHISPERS.2019.8921273.

[9] K. Feng, Y. Zhao, J. C.-W. Chan, S. G. Kong, X. Zhang, and B. Wang, "Mosaic Convolution-Attention Network for Demosaicing Multispectral Filter Array Images," *IEEE Transactions on Computational Imaging*, vol. 7, pp. 864–878, 2021, doi: 10.1109/TCI.2021.3102052.

[10] K. Shinoda, S. Yoshiba, and M. Hasegawa, "Deep demosaicking for multispectral filter arrays," *arXiv:1808.08021 [eess]*, Oct. 2018,

Accessed: Feb. 18, 2021. [Online]. Available: http://arxiv.org/abs/1808.08021

[11] M. Gupta, V. Rathi, and P. Goyal, "Adaptive and Progressive Multispectral Image Demosaicking," *IEEE Transactions on Computational Imaging*, vol. 8, pp. 69–80, 2022.

[12] J. Mizutani, S. Ogawa, K. Shinoda, M. Hasegawa, and S. Kato, "Multispectral demosaicking algorithm based on inter-channel correlation," in *2014 IEEE Visual Communications and Image Processing Conference*, Dec. 2014, pp. 474–477. doi: 10.1109/VCIP.2014.7051609.

[13] S. Mihoubi, O. Losson, B. Mathon, and L. Macaire, "Multispectral Demosaicing Using Pseudo-Panchromatic Image," *IEEE Trans. Comput. Imaging*, vol. 3, no. 4, pp. 982–995, Dec. 2017, doi: 10.1109/TCI.2017.2691553.

[14] S. Ogawa *et al.*, "Demosaicking Method for Multispectral Images Based on Spatial Gradient and Inter-channel Correlation," in *Image and Signal Processing*, Cham, 2016, pp. 157–166. doi: 10.1007/978-3-319-33618-3_17.

[15] V. Rathi and P. Goyal, "Generic Multispectral Image Demosaicking Algorithm and New Performance Evaluation Metric," in *Computer Vision and Image Processing*, Cham, 2022, pp. 45–57. doi: 10.1007/978-3-031-11346-8_5.

[16] J. Brauers and T. Aach, "A color filter array based multispectral camera," in *12. Workshop Farbbildverarbeitung*, 2006.

[17] G. Tsagkatakis, M. Bloemen, B. Geelen, M. Jayapala, and P. Tsakalides, "Graph and Rank Regularized Matrix Recovery for Snapshot Spectral Image Demosaicing," *IEEE Trans. Comput. Imaging*, vol. 5, no. 2, pp. 301–316, Jun. 2019, doi: 10.1109/TCI.2018.2888989.

[18] L. Bian, Y. Wang, and J. Zhang, "Generalized MSFA Engineering With Structural and Adaptive Nonlocal Demosaicing," *IEEE Transactions on Image Processing*, vol. 30, pp. 7867–7877, 2021, doi: 10.1109/TIP.2021.3108913.

[19] M. Kawase, K. Shinoda, and M. Hasegawa, "Demosaicking Using a Spatial Reference Image for an Anti-Aliasing Multispectral Filter Array," *IEEE Trans. on Image Process.*, vol. 28, no. 10, pp. 4984–4996, Oct. 2019, doi: 10.1109/TIP.2019.2910392.

[20] K. Dijkstra, J. van de Loosdrecht, L. R. B. Schomaker, and M. A. Wiering, "Hyperspectral demosaicking and crosstalk correction using deep learning," *Machine Vision and Applications*, vol. 30, no. 1, pp. 1–21, Feb. 2019, doi: 10.1007/s00138-018-0965-4.

[21] B. Arad *et al.*, "NTIRE 2022 Spectral Demosaicing Challenge and Data Set," in *Proceedings of the IEEE/CVF Conference on Computer Vision and Pattern Recognition*, 2022, pp. 882–896.

[22] S. Liu, Y. Zhang, J. Chen, L. K. Pang, and S. Rahardja, "A Deep Joint Network for Multispectral Demosaicking Based on Pseudo-Panchromatic Images," *IEEE Journal of Selected Topics in Signal Processing*, pp. 1–1, 2022, doi: 10.1109/JSTSP.2022.3172865.

[23] P. Li *et al.*, "Deep learning approach for hyperspectral image demosaicking, spectral correction and high-resolution RGB reconstruction," *Computer Methods in Biomechanics and Biomedical Engineering: Imaging & Visualization*, vol. 10, no. 4, pp. 409–417, Jul. 2022, doi: 10.1080/21681163.2021.1997646.

[24] S. P. Jaiswal, L. Fang, V. Jakhetiya, J. Pang, K. Mueller, and O. C. Au, "Adaptive Multispectral Demosaicking Based on Frequency-Domain Analysis of Spectral Correlation," *IEEE Transactions on Image Processing*, vol. 26, no. 2, pp. 953–968, Feb. 2017, doi: 10.1109/TIP.2016.2634120.

[25] S. Mihoubi, "Snapshot multispectral image demosaicing and classification," Theses, Université de Lille, 2018. Accessed: Oct. 16, 2022. [Online]. Available: https://hal.archives-ouvertes.fr/tel-01953493

[26] D. Chen, J. Tachella, and M. E. Davies, "Equivariant Imaging: Learning Beyond the Range Space," presented at the Proceedings of the IEEE/CVF International Conference on Computer Vision, 2021, pp. 4379–4388. Accessed: Nov. 10, 2022. [Online]. Available: https://openaccess.thecvf.com/content/ICCV2021/html/Chen_Equivariant_Imaging_Learning_Beyond_the_Range_Space_ICCV_2021_paper.html

[27] Y. Niu, J. Ouyang, W. Zuo, and F. Wang, "Low Cost Edge Sensing for High Quality Demosaicking," *IEEE Transactions on Image Processing*, vol. 28, no. 5, pp. 2415–2427, May 2019, doi: 10.1109/TIP.2018.2883815.

[28] F. Iandola and K. Keutzer, "Keynote: small neural nets are beautiful: enabling embedded systems with small deep-neural- network architectures," in *2017 International Conference on Hardware/Software Codesign and System Synthesis (CODES+ISSS)*, 2017, pp. 1–10. doi: 10.1145/3125502.3125606.

[29] G. Qian, J. Gu, J. S. Ren, C. Dong, F. Zhao, and J. Lin, "Trinity of Pixel Enhancement: a Joint Solution for Demosaicking, Denoising and Super-Resolution," *arXiv:1905.02538 [cs, eess]*, May 2019, Accessed: Dec. 25, 2020. [Online]. Available: http://arxiv.org/abs/1905.02538

[30] E. L. Wisotzky, C. Daudkane, A. Hilsmann, and P. Eisert, "Hyperspectral Demosaicing of Snapshot Camera Images Using Deep Learning," in *Pattern Recognition*, Cham, 2022, pp. 198–212. doi: 10.1007/978-3-031-16788-1_13.

[31] A. Bruhn, J. Weickert, and C. Schnörr, "Lucas/Kanade Meets Horn/Schunck: Combining Local and Global Optic Flow Methods," *International Journal of Computer Vision*, vol. 61, no. 3, pp. 211–231, Feb. 2005, doi: 10.1023/B:VISI.0000045324.43199.43.

[32] W. Shi *et al.*, "Real-time single image and video super-resolution using an efficient sub-pixel convolutional neural network," in *Proceedings of the IEEE conference on computer vision and pattern recognition*, 2016, pp. 1874–1883.

[33] J. Hu, L. Shen, and G. Sun, "Squeeze-and-Excitation Networks," p. 10.

[34] B. Arad and O. Ben-Shahar, "Sparse Recovery of Hyperspectral Signal from Natural RGB Images," in *Computer Vision – ECCV 2016*, Cham, 2016, pp. 19–34. doi: 10.1007/978-3-319-46478-7_2.

[35] Z. Wang, A. C. Bovik, H. R. Sheikh, and E. P. Simoncelli, "Image quality assessment: from error visibility to structural similarity," *IEEE Transactions on Image Processing*, vol. 13, no. 4, pp. 600–612, Apr. 2004, doi: 10.1109/TIP.2003.819861.

[36] F. A. Kruse *et al.*, "The spectral image processing system (SIPS)—interactive visualization and analysis of imaging spectrometer data," *Remote Sensing of Environment*, vol. 44, no. 2, pp. 145–163, May 1993, doi: 10.1016/0034-4257(93)90013-N.

[37] L. Wald, "Quality of high resolution synthesised images: Is there a simple criterion?," in *Third conference "Fusion of Earth data: merging point measurements, raster maps and remotely sensed images"*, Sophia Antipolis, France, 2000, pp. 99–103. Accessed: Aug. 23, 2022. [Online]. Available: https://hal.archives-ouvertes.fr/hal-00395027

[38] J. Yang, Y.-Q. Zhao, C. Yi, and J. C.-W. Chan, "No-reference hyperspectral image quality assessment via quality-sensitive features learning," *Remote Sensing*, vol. 9, no. 4, p. 305, 2017.

[39] D. Chen, J. Tachella, and M. E. Davies, "Robust Equivariant Imaging: a fully unsupervised framework for learning to image from noisy and partial measurements," *arXiv:2111.12855 [cs, eess]*, Nov. 2021, Accessed: Dec. 13, 2021. [Online]. Available: http://arxiv.org/abs/2111.12855

[40] S. Guo, Z. Liang, and L. Zhang, "Joint Denoising and Demosaicking With Green Channel Prior for Real-World Burst Images," *IEEE Transactions on Image Processing*, vol. 30, pp. 6930–6942, 2021, doi: 10.1109/TIP.2021.3100312.

[41] N. Li, B. L. Teurnier, M. Boffety, F. Goudail, Y. Zhao, and Q. Pan, "No-Reference Physics-Based Quality Assessment of Polarization Images and Its Application to Demosaicking," *IEEE Transactions on Image Processing*, vol. 30, pp. 8983–8998, 2021, doi: 10.1109/TIP.2021.3122085.

[42] B. Wang, C. Lu, D. Yan, and Y. Zhao, "Learning Pixel-Adaptive Weights for Portrait Photo Retouching." arXiv, Dec. 07, 2021. doi: 10.48550/arXiv.2112.03536.

[43] B. Wang, Y. Zhao, and X. Li, "Multiple Instance Graph Learning for Weakly Supervised Remote Sensing Object Detection," *IEEE Transactions on Geoscience and Remote Sensing*, vol. 60, pp. 1–12, 2022, doi: 10.1109/TGRS.2021.3123231.

[44] W. Zhou, L. Zhang, S. Gao, and X. Lou, "Gradient-Based Feature Extraction From Raw Bayer Pattern Images," *IEEE Transactions on Image Processing*, vol. 30, pp. 5122–5137, 2021, doi: 10.1109/TIP.2021.3067166.